\documentclass[aps,prl,showpacs,twocolumn]{revtex4-1}

\usepackage{graphicx,bbm}

\def\ave#1{\langle #1 \rangle}

\def\tit#1{}

\begin{document}

\title{Time irreversible billiards with piecewise-straight trajectories}

 \author{Giulio Casati}
\email[]{giulio.casati@uninsubria.it} \affiliation{Center for
Nonlinear and Complex Systems, Universit\`a degli Studi
dell'Insubria, Como, Italy} \affiliation{CNR-INFM and Istituto
Nazionale di Fisica Nucleare, Sezione di Milano, Milan, Italy }
\author{Toma\v z Prosen}
\email[]{tomaz.prosem@fmf.uni-lj.si}\affiliation{Physics department, Faculty of Mathematics and Physics,
University of Ljubljana, Ljubljana, Slovenia}

\date{\today}

\begin{abstract}
We introduce a new family of billiards which break time reversal symmetry in spite of having piece-wise straight trajectories. We show that our billiards preserve the ergodic and mixing properties of conventional billiards while they may turn into exponential the power law decay of correlations characteristic of Sinai type billiards. Such billiards can be implemented by squeezing the transverse magnetic field along lines or along one-dimensional manifolds.
\end{abstract}



\maketitle

{\em Introduction.-} Billiards are mathematical models where one
or more particles move freely in a container and collide with its walls and/or with each
other~\cite{Sinai,Dettmann}. The dynamical properties of such models,   which play a basic role in the understanding of a large variety of physical phenomena,  are determined by the shape of the
walls of the container. Their behavior is very rich and can range  from completely integrable, regular motion, via generic mixed phase space dynamics, to
fully chaotic motion. 
Classical and quantum billiards play a basic role in nonlinear physics since in several cases they allow a rigorous theoretical analysis and also because they are very convenient for numerical simulation. They have been used in relation to the celebrated Boltzmann hypothesis in statistical mechanics, the Riemann hypothesis, number theory, the Lorentz gas, introduced in 1905 to describe electricity, and finally in quantum chaos when solving the
Helmholtz equation with Dirichlet or Neumann boundary
conditions. They have been fundamental in the connection between classical chaotic motion and the statistical properties of eigenvalues and eigenfunctions of the corresponding quantum system.

Billiard systems are of significant
interest also in experimental physics and in applications
related to low-dimensional nanostructures.
Recent developments in nanotechnology have made it possible to experimentally
realize such systems by electrostatically confining a
two-dimensional electron gas  in high mobility heterostructures.
Billiards were experimentally realized first
in flat microwave resonators in the early 1990's \cite{Stoeckmann,Sridhar} and soon
later in semiconductor nano-structures such as quantum dots \cite{Marcus}
which  are studied in view of emerging applications, including quantum
computation. They exhibit deterministic ballistic motion
of the electrons and provide the possibility
to tune their shape, size, and electron number.
Billiards have furthermore been experimentally realized for investigations in a variety of fields:
in room acoustics and elastomechanics of solid plates, in atom optics, where
ultracold atoms reflect from laser beams, in optics of dielectric
microresonators, etc. Understanding details of classical billiard dynamics even had a very practical application in controlling directed emission in solid-state microlasers \cite{Stone}.
 
From the theoretical point of view and for physical applications it is important to have billiards with broken time reversal symmetry. There are no example of such models so far. The closest models are billiards in perpendicular magnetic field which have been shown to exhibit quantum interference effects like weak localization, Altshuler-Aronov-Spivak oscillations, and conductance fluctuations~\cite{nakamura}.
 However dynamical properties of such models are much more complicated. In particular, with the exception of billiards with strictly convex boundaries~\cite{gutkin}, the application of a uniform magnetic field to an ergodic billiard leads to appearance of islands of stability in phase space and to a very complicated dynamics.
Indeed the presence of the magnetic field changes the
stability properties of the billiard motion. The Jacobian matrix
of the motion between collisions, which is parabolic in the case without magnetic field, becomes elliptic. Moreover the magnetic filed changes also the parabolic matrix corresponding to boundary collisions. As a result, the study of the stability properties of the orbits becomes very difficult \cite{robnik}.

In this paper we introduce a new class of billiards -- i.e. conservative (Hamiltonian) systems in which motion takes place on straight lines apart from collisions with the boundary -- which are time reversible symmetry breaking and do not alter the ergodic and mixing property of conventional billiards. It is remarkable that the decay of correlations can be turned from power law to exponential by just tuning the intensity of the magnetic field. 
These billiards may become important tools in investigation in several fields of nanoscience.

\begin{figure}[t!]
\centerline{\includegraphics[width=0.48\textwidth]{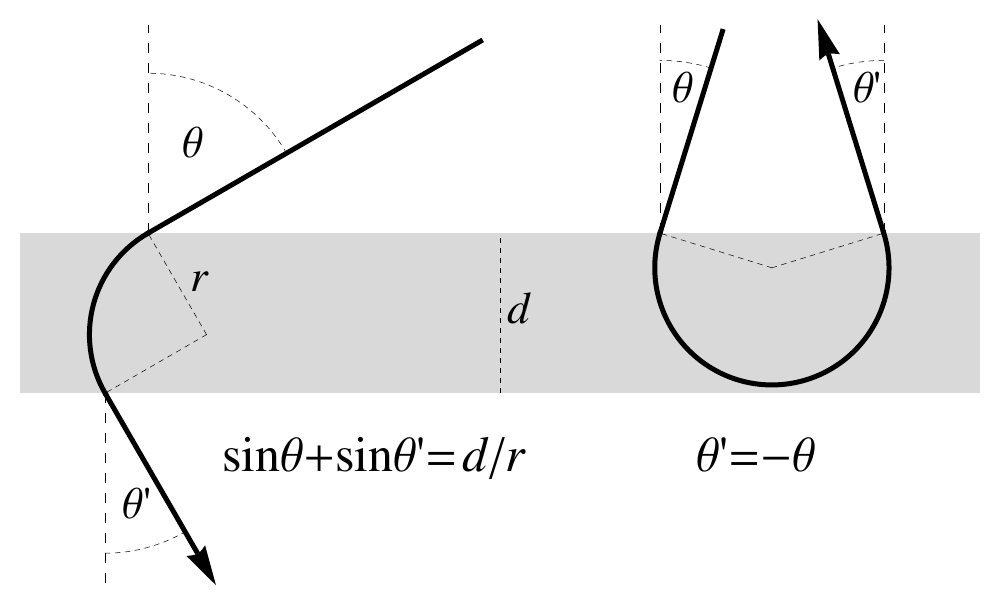}}
\caption{Schematic illustration of time-irreversible collision rules. The light-grey region of thickness $d$ is considered to be in a transverse magnetic field where the Larmour radius is $r$. At the end we consider the limits
$d\to 0$ and $r\to 0$, such that the ratio $\eta=d/r$ is finite. On the left we show penetration which is possible when $|d/r-\sin \theta| < 1$, while in the opposite case we have a specular reflection (shown on the right).}
\label{fig:scheme}
\end{figure}

{\em Billiards with magnetic boundaries.-}
The key ingredient in our idea is a one dimensional magnetic manifold, a strip of a transverse magnetic field which we shall eventually think of squeezing into a line.
Imagine  such a locally flat strip of width $d$, onto which a charged particle impacts under angle $\theta$ (see Fig.~\ref{fig:scheme}). Then, depending on the strength of the magnetic field $B$ in the strip, yielding a Larmour radius $r=B^{-1}$ (we shall use units where the particle mass, charge, and velocity $m=e=v=1$), the particle either penetrates the magnetic strip, if $|\sin\theta - d/r| < 1$, or oppositely, is reflected. Without loosing generality we may assume that the magnetic field is pointing upward (from the plane of the figure), i.e. $B>0$.
Introducing the non-dimensional magnetic parameter 
\begin{equation}
\eta := \frac{d}{r} = B d,
\label{eq:eta}
\end{equation}
one sees that the angle of incidence $\theta$ and the angle of emittance $\theta'$ are related by a trivial trigonometric identity (see Fig.~\ref{fig:scheme}):
\begin{equation}
\sin\theta+\sin\theta' = \eta.
\end{equation}
This geometric relation remains valid in the limit where the magnetic strip becomes infinitely narrow $d\to 0$ if at the same time the strength of magnetic field is increased
$B\to\infty$, such that $\eta$ remains finite.
The penetration/collision rule for such magnetic boundary then becomes
\begin{equation}
\theta' = \left\{\begin{array}{ll}
\arcsin(\eta -\sin\theta), &\mbox{if } |\eta -\sin\theta| \le 1 ;  \\
-\theta, &\mbox{if } |\eta - \sin\theta| > 1.
\end{array}\right.
\end{equation}
 Notice that, in spite of the fact that motion always occurs along straight lines, the magnetic boundary manifestly breaks time reversal symmetry: the time-reversed trajectory is in general not a valid trajectory.  The non trivial range of magnetic parameter is $0<\eta<2$.
  For $\eta \ge 2$, all trajectories get specularly reflected as in conventional billiards, while for $\eta=0$ the walls are simply transparent.

\begin{figure}[t!]
\centerline{\includegraphics[width=0.48\textwidth]{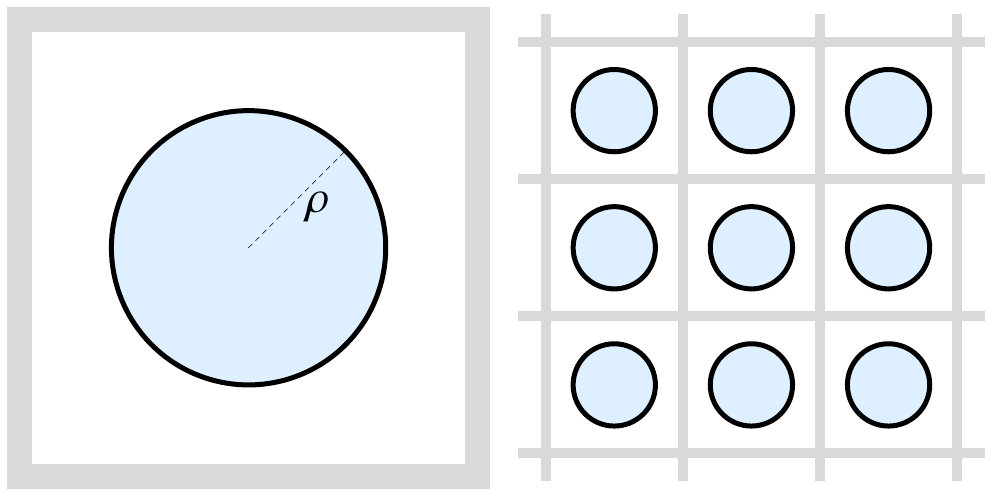}}
\caption{(Color online)~Schematic picture of time-irreversible Sinai billiard. The boundary of a unit square (in light gray) is considered to be magnetic, while the circular obstacle of radius $\rho$ is considered to be a hard-wall boundary. When the particle penetrates through the magnetic boundary we consider periodic boundary condition and inject it on the opposite side of the square. Alternatively, we may consider a magnetic heterostructure (shown in the right) where magnetic field is squeezed onto a two-dimensional cartesian net.}
\label{fig:billiard}
\end{figure}

Considering orbit stability, the magnetic boundary behaves in exactly the same way as a hard-wall boundary. In particular, if the magnetic boundary is locally flat, it maps parallel trajectories to parallel trajectories. More generally, it preserves the spread of the sines-of-the-angle of a beam of trajectories, since ${\rm d} \sin\theta=-{\rm d}\sin\theta'$ where ${\rm d}$ denotes the differential. If one uses appropriate canonical Birkhoff coordinates for the bounce map one can thus show that flat magnetic boundaries  can not change the stability focusing/defocusing property of the billiard. Since the corresponding Poincar\' e map is symplectic the uniform measure is an invariant measure of the dynamics. 
 As a generic example, we study in the following the Sinai billiard with magnetic outer-boundary as indicated in Fig.~\ref{fig:billiard}. As the particles will in general be able to penetrate through the outer boundary (a unit square), we have to impose periodic boundary conditions in order to keep the phase space finite. However, the dynamics of such a billiard on the unfolded configuration space may be interpreted as the dynamics of charged particles in a magnetic heterostructure where the magnetic field is squeezed onto a Cartesian net with circular hard-wall obstacles of radius $\rho$ placed in the center of each cell defined by the net.

\begin{figure}[t!]
\centerline{\includegraphics[width=0.48\textwidth]{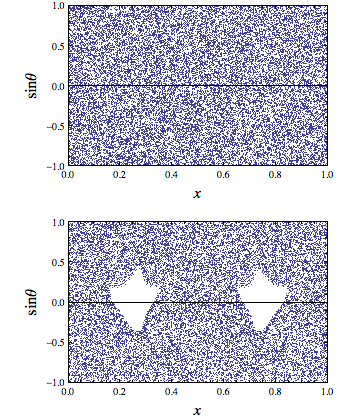}}
\caption{Poincar\' e surface of section for the Sinai Billiard with magnetic boundary ($\eta=0.5$) [top panel] and for the Sinai Billard with all-hard-wall-boundary and a uniform magnetic field with Larmour radius $r=0.5$ [bottom panel]. In both cases the discs radius is $\rho=0.3$.
We note the appearance of islands of regular motion in the second case, while in the first case the motion is ergodic.
Here we use only one side of the unit square as a surface of section ($0\le x\le 1$) with the canonically conjugate momentum coordinate $\sin\theta$.}
\label{fig:sos}
\end{figure}

\begin{figure}[t!]
\centerline{\includegraphics[width=0.5\textwidth]{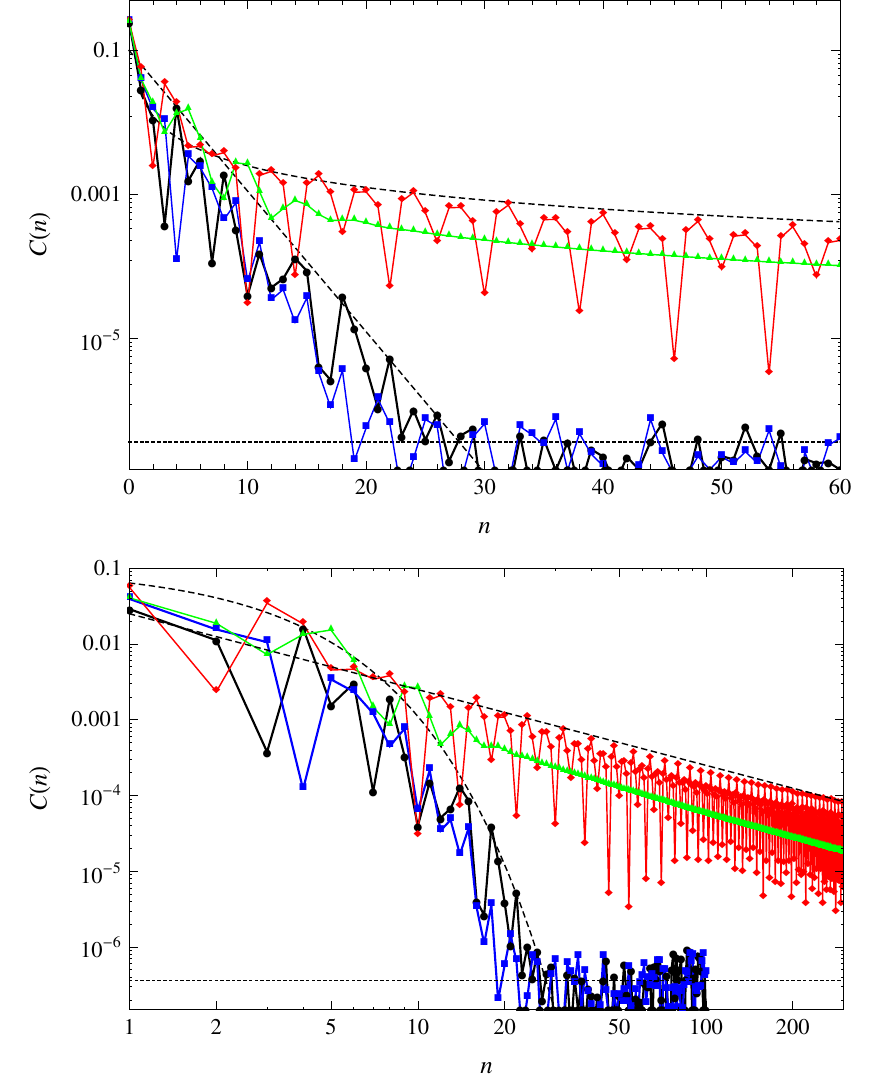}}
\caption{(Color online)~Velocity-velocity correlation function $C(n)=\ave{v_x(0)v_x(n)}$ where $n$ counts the number of collisions/penetrations with/through the magnetic boundary. We take $\rho=0.4$, and consider different values of
magnetic parameter $\eta=(\sqrt{5}-1)/2$ (black circles), $\eta=0.1$ (blue squares),  for which we find exponential decay $\propto \exp(-0.45 n)$ (indicated by a dashed curve), and
again $\rho=0.4$ with $\eta=0$ (green triangles), and $\rho=0.3$ with $\eta=(\sqrt{5}-1)/2$ (red diamonds), for which we find $1/n$ decay (indicated by another dashed curve).
The upper panel shows data in semi-log, while the lower panel in log-log scale.}
\label{fig:corr}
\end{figure}

As mentioned above, the standard proof of ergodicity and mixing for dispersing (Sinai-type) billiards \cite{Sinai}, goes through along the same lines if pieces of flat boundary are replaced by our magnetic boundary. Thus our billiard should be ergodic and mixing with positive Lyapounov exponent independently on the value of $\eta$. In Fig.~\ref{fig:sos} we plot the Poincar\' e surface of section, with respect to one side of the square (which is arbitrary due to the $C_4$ symmetry) of a Sinai billiard with magnetic boundary, and compare it to a surface of section of the same Sinai billiard with a uniform magnetic field of comparable `average strength'. We note that, as expected, in the second case there are islands of regular motion in sharp contrast to the ergodic behavior of our magnetic billiard. For the latter we have made careful numerical investigations and all evidence show that there are no islands. In Fig.~\ref{fig:sos} we just show one example to compare with the case of a uniform magnetic field.

The time-irreversible nature of our billiard leads to the appearance of interesting dynamical properties.  For example, a characteristic feature of Sinai like billiards is the presence of the so called long time tails. These tails are caused by the existence of a continuous family -- of zero measure -- of marginally stable periodic orbits which lead to power law decay of correlations in spite of the mixing and exponentially unstable motion. Our magnetic billiard exhibits a richer behavior. In particular, by changing the intensity of the applied field one can control the decay of correlations without altering the general ergodic properties of the system. Indeed, upon changing $\eta$, marginally stable periodic orbits may disappear thus restoring the exponential decay of correlations.
However, for $\eta \ge 1$, marginally stable periodic orbits do exist and are the same ``bouncing balls" orbits ($\theta=0$, $\theta= \pi/2$) of the conventional Sinai billiard. Such orbits are also present for $\eta <1$ provided the disc radius $\rho < \rho_{cr}= \sqrt{2}/4$. This restriction in the disc radius is necessary to allow the presence of diagonal (`diamond' shaped) marginally stable periodic orbits ($\theta= \pi/4$) connecting the mid points of the squares sides. However, for $0 < \eta <1$  and for $\rho > \rho_{cr}$,  no marginally stable periodic orbits exists and therefore one expects exponential decay of correlations. These expectations are corroborated by numerical results shown in Fig.~4 in semi-log and log-log scale. Here we consider the Poincare map (with respect to all 4-sides of the square, or with respect to the magnetic net of the 2D periodic heterostructure), and compute the velocity-velocity autocorrelation function
\begin{equation}
C(n) = \ave{v_x(0)v_x(n)} = \lim_{M\to\infty}\frac{1}{M} \sum_{m=1}^M v_x(m)v_x(m+n)
\end{equation}
We note that $\ave{v_x} = 0$, so the decay of $C(n)$ is an indicator of mixing.
 It is seen that, as expected, for $\rho= 0.3 <\rho_{cr}$ the correlation function $C(n)$ decays algebraically, as $\sim 1/n$. Quite obviously, the same algebraic decay takes place for the case with no magnetic field ($\eta=0$, i.e. infinite-horizon Lorentz gas) independently on the disc radius. Instead for  $\rho= 0.4 >\rho_{cr}$ and for $\eta < 1$, the
 decay of $C(n)$ is clearly exponential $|C(n)| \sim \exp(-\nu n)$. 
 
 The rate of exponential decay $\nu \approx 0.45$ appear not to sensibly depend on the
 magnetic field parameter $\eta$. Analogously, the Lyapounov exponent, which is around $\lambda\approx 1$, only slightly depends on the value of $\eta$.
 
 We note that in our example of modified Sinai billiard, just like in the usual Sinai billiard, the central disk is relevant for introducing exponential instability into orbits' dynamics. We have also studied a similar magnetic-mesh heterostructure without circular obstacles. Even though such a dynamics cannot produce Lyapunov instability it still makes dynamics non-separable and non-integrable for typical values of parameter $\eta$. 
On the basis of numerical results we argue that such dynamics should be qualitatively similar to the dynamics of polygonal billiards. In particular, for our cartesian magnetic mesh the rate of ergodicity seems very similar to the triangular billiards with two angles being irrational multiples of $\pi$ while the third angle is a rational multiple of $\pi$ \cite{CasPro}. Indeed, our magnetic periodic boundary conditions produce discontinuities in the Poincar\' e map -- namely
trajectories which are only slightly different can either penetrate or reflect from the boundary. Similar discontinuities are produced by scattering from non-right angles in polygonal billiards.

{\em Conclusion.-} we believe we provided the first example of a billiard type dynamical system with (piece-wise) straight trajectories which breaks time-reversal invariance and is ergodic and mixing with possibly exponential decay of correlations.
The breaking of time reversal symmetry is known to have important consequences in several field of physics. One important example is the generation of directed transport. Very recently it has been shown that it can play a crucial role
in the design of high performance thermoelectric devices for efficient power generation and refrigeration~\cite{casati}. In this relation, time irreversible billiards are the natural candidates to provide a better understanding of the dynamical mechanisms which determine the fundamental limits that thermodynamic imposes on thermal machines, a central issue in physics and of increasing relevance in the future society.

We acknowledge useful discussion with Carl Dettmann. The work has been supported by the grants P1-044 and J1-2208 of Slovenian Research Agency (ARRS), MIUR-PRIN 2008 and by Regione Lombardia.

\end{document}